\documentclass[preprint,aps,prb,amsmath,amssymb]{revtex4-1}
\usepackage{amsmath}
\usepackage{amsfonts}
\usepackage{amssymb}
\usepackage{graphics}
\usepackage{float}
\usepackage{graphicx}
\usepackage{epstopdf}
\usepackage[compact]{titlesec}
\usepackage{natbib}
\usepackage{threeparttable}
\usepackage[titletoc,title]{appendix}
\usepackage{lipsum}
\usepackage{varwidth}
\usepackage{tabularx}
\usepackage{color}

\begin{document}
\title{Spin-dependent magneto-thermopower of narrow-gap lead chalcogenide quantum wells}
\author{Parijat Sengupta}
\affiliation{Dept. of Electrical and Computer Engineering, University of Illinois, Chicago, IL 60607.}
\author{Junxia Shi}
\affiliation{Dept. of Electrical and Computer Engineering, University of Illinois, Chicago, IL 60607.}

\begin{abstract}
{A semi-classical analysis of magneto-thermopower behaviour, namely, the \textit{Seebeck} and \textit{Nernst} effect (NE) in quantum wells of IV-VI lead salts with significant extrinsic Rashba spin-orbit coupling (RSOC) is performed in this report. In addition to the spin-dependent Seebeck effect that has been observed before, we also theoretically predict a similar spin-delineated behaviour for its thermal analog, the spin-dependent NE. The choice of lead salts follows from a two-fold advantage they offer, in part, to their superior thermoelectric properties, especially PbTe, while their low band gaps and high spin-orbit coupling make them ideal candidates to study \textit{RSOC} in nanostructures. The calculations show a larger longitudinal magneto-thermopower for the spin-up electrons while the transverse components are nearly identical. In contrast, for a magnetic field free case, the related power factor calculations reveal a significantly higher contribution from the spin-down ensemble and suffer a reduction with an increase in the electron density. We also discuss qualitatively the limitations of the semi-classical approach for the extreme case of a high magnetic field and allude to the observed thermopower behaviour when the quantum Hall regime is operational. Finally, techniques to modulate the thermopower are briefly outlined.}
\end{abstract}
\maketitle

\section*{Introduction}
\begin{figure}[!t]
\centering
\includegraphics[scale=0.9]{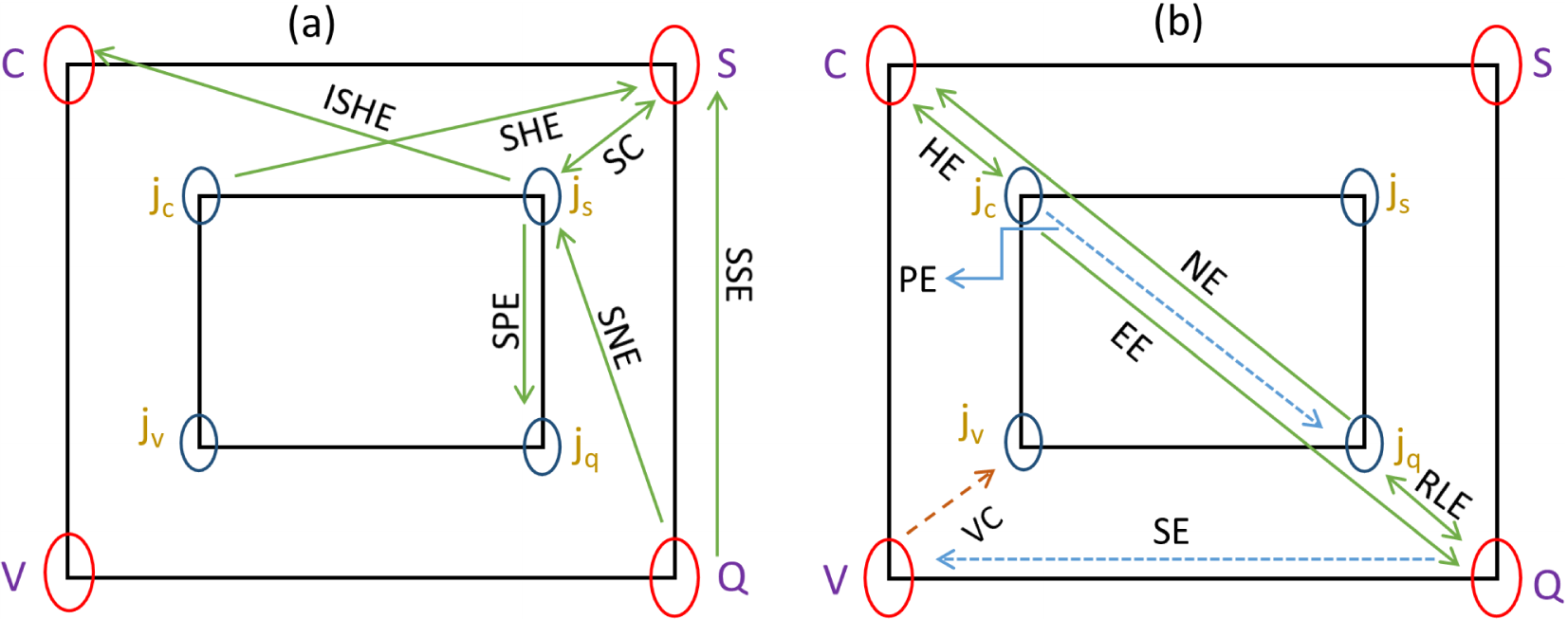}
\caption{The four corners of the outer and embedded inner squares in the two-paneled figure are symbolically marked by specific gradients (potential, in case of charge (C) and spin (S) and temperature for a thermal (Q) bias) and their current counterparts in presence of a magnetic field. The arrow-denoted linkages show the several processes that may arise appearing as flow of charge ($ j_{C} $), spin ( $ j_{S} $), or thermal ($ j_{Q} $) currents. Some processes lead to the creation of a potential or temperature gradient. The left figure (a) sketches the spin-driven transport processes; for instance, the spin Seebeck effect (\textbf{SSE}) is the generation of a spin voltage by a temperature gradient while the reciprocal spin Peltier effect (\textbf{SPE}) indicates a finite heat current arising from a spin injection. Similarly, the spin Nernst effect (\textbf{SNE}) describes a transverse spin current created on account of a longitudinal temperature gradient. The spin Hall effect (\textbf{SHE}) is the production of a finite spin current density that flows perpendicular to a charge current. Inverse SHE (\textbf{ISHE}) is the reciprocal process of SHE. The spin conductivity (\textbf{SC}) is spin current that flows for a finite spin accumulation, which in the cases listed here are controlled by a thermal gradient. The more conventional effects, such as Nernst (\textbf{NE}), Ettingshausen (\textbf{EE}) Righi-Leduc (\textbf{RLE}), Hall (\textbf{HE}) are observed for a finite $ \mathbf{H} $ (the right panel (b) illustrates these collective effects) and constitute classical magneto-thermoelectrics. The Seebeck effect (SE) and its reciprocal, the Peltier process illustrate the interplay between charge and heat currents when the magnetic field is absent. The Nernst and Seebeck effects are described in this work; a quantitative determination is made by connecting the potential and temperature gradient via a tensor matrix. Lastly, the valley (V) is an additional degree of freedom and serves as the origin of an anomalous current ($ j_{V} $) and the valley Hall effect (VHE) for a finite Berry curvature. The Berry curvature is the momentum-space analogue of a real-space magnetic field. A difference in carrier population reflects as $ j_{v} $, the valley voltage.}
\label{schema}
\vspace{-0.3cm}
\end{figure}

Energy transport processes primarily aimed at efficient power generation, transmission, and inter-conversion between diverse forms have been tested on multiple functional materials~\cite{zhao2014review,elsheikh2014review}. This brings within its ambit processes (both reversible and otherwise) reliant on coupled thermal and electric phenomena. There has been significant progress achieved in our understanding of distinct modes of energy transport, in particular, processes that conjunct the charge and spin of carriers in a solid~\cite{brown2008improved,ozaeta2014predicted}. There also exist several illustrations of intriguing charge and spin excitations in systems under an external magnetic field, which, when driven into non-equilibrium by an applied temperature and voltage gradient manifest as a current (charge/spin/thermal) or an open-circuit electromotive force. A great variety of such processes can appear, for example, in the magnon-dependent spin-Seebeck effect, observed as a spin-voltage in response to a thermal gradient in a ferromagnet~\cite{uchida2008observation,xiao2010theory}. More such processes are pictorially summarized in Fig.~\ref{schema} where each named-pathway describes a microscopic response to a predefined non-equilibrium situation. From a more fundamental standpoint, heat currents through galvano- and thermo-magnetic effects can reveal the interaction of the carriers with the lattice. While ordinarily the magneto-thermopower connected phenomena observed in metals and conventional semiconductors is not significant, certain compounds, for instance, Bi$_{0.91}$Sb$_{0.09}$ in a magnetic field of 0.5 T shows a two-fold increase~\cite{ovsyannikov2009high} in \textit{ZT}. Likewise, the thermopower of PbTe, a well-known thermoelectric, exhibits a significant variation in an external magnetic field under pressure while a large improvement was also confirmed when alloyed~\cite{osinniy2005pb} with the magnetic dopant, Mn, to form Pb$_{x}$Mn$_{1-x}$Te. In addition, the presence of non-trivial electronic states can lead to a significant increment in the magneto-thermopower, specific illustrations of which can be found in large magneto-resistance effect in gapless Ag$_{2}$Se and Ag$_{2}$Te composites with Dirac (linear) dispersion. A more robust evidence of the role of Dirac carriers was also uncovered as a giant magneto-thermopower in bulk Dirac materials such as (Sn/Cr)MnBi$_{2}$~\cite{wang2012large}. Generally, in the presence of a magnetic field that can lead to formation of Landau levels, the thermopower tensor reveals important characteristic features such as quantum lifetime, disorder potential, and electrical transport coefficients including carrier mobility and the effective mass. From an application perspective, the discovery of newer materials and the opportunity afforded by recently established (and better-understood) energy flow mechanisms complement the primary goal of designing thermoelectric (\textit{TE}) devices that seek an optimal conversion (embodied in an enhanced thermoelectric figure of merit, \textit{ZT}) of heat to charge and spin currents.  

While a large set of thermoelectric data exists for bulk materials, dimensionally-confined structures that span the whole gamut from quantum wells to dots including those derived from nanocrystals~\cite{schaller2004high} widen the scope of applicability and expand the accompanying thermal device design space. It is now verifiably proven~\cite{mahan1996best} that a large \textit{ZT} ensues from quantization of electron states through dimensional reduction, for instance, the thermoelectrics of nanostructured quantum wells and thin films report a marked improvement over their bulk counterparts. Thermoelectric devices can attain desired efficiency by adjustments to the thermal and electric conductivity, a pair of transport numbers with a marked degree of reliance on quantum confinement, where they are amenable to further changes via magnetic fields, mechanical strain, pressure, and nature of added dopants. In this report, we set up a basic formalism for spin-dependent magneto-\textit{Seebeck} and \textit{Nernst-Ettingshausen} effect, sometimes simply referred to as the \textit{Nernst} effect) and apply them to the lead chalcogenide, PbTe, a well-researched thermoelectric. These magneto-thermopower calculations with the correct set of material parameters can be extended to PbS and PbSe, the two other important members of the lead chalcogenide family. The spin-dependent behaviour of magneto-thermopower (later we also extract from the analytic calculations the zero magnetic case for simple thermopower) is essentially a manifestation of the spin splitting introduced by the the Rashba coupling dominant in these lead salts. The electronic spin degree of freedom in compounds with a large spin-orbit coupling often stamps its imprint by spin-polarizing the carriers and breaking degeneracy, which in thin films of Pb-salts can be decidedly strong through the extrinsic Rashba spin-orbit coupling (\textit{RSOC}). The extrinsic \textit{RSOC} is pronounced~\cite{winkler2003spin} in lead salts, particularly in narrow-gap PbTe $\left(0.19\, eV\right)$ and receives a further boost from the substantial intrinsic spin-orbit coupling $\left(0.77\, eV\right)$; this, as we qualitatively show later, imparts a marked spin-dependent character to the magneto-\textit{Seebeck} and \textit{Nernst-Ettingshausen} effects. For simplicity and to better understand the behaviour of the magneto-thermopower, we carry out calculations in a classical framework and ignore any Landau level (LL) quantization; their relevance though, is qualitatively discussed in context of the quantum Hall regime and the magneto-thermopower behaviour thereof. The thermopower which is tensorial in presence of a magnetic field is obtained by a direct application of Mott's formula.~\cite{lunde2005mott}. Further, as thermal effects are crucially contingent on the nature of energy bands, we set up (and numerically solve) a Kane-like \textit{k.p} Hamiltonian (adapted from the bulk version, which is $ 4 \times 4 $) that describes the dispersion around the high-symmetry \textit{L}-valleys (four in number) for the lead chalcogenide quantum wells.~\cite{dimmock1964band,kang1997electronic}

The results presented in this report draw attention to the magnetic field alteration of the thermopower and quantitatively predict its longitudinal and transverse components for conduction electrons located in the vicinity of the $\left[111\right] $ parallel \textit{L}-valley in PbTe thin films. The calculations are spin-resolved, each spin-ensemble created by the strong \textit{RSOC} in PbTe films. The role of \textit{RSOC}through spin splitting manifests as a larger longitudinal magneto-thermopower for the spin-up electrons while its transverse counterpart does not show any discernible difference. The derived results, beyond the predicted \textit{RSOC}-governed variance, reveal a direct relationship to the dominant energy-dependent scattering mechanisms operational in the film. Remarkably though, as we show, for processes that are energy independent, the transverse part of thermopower vanishes even for a finite magnetic field. We also point out that a similar cessation of the transverse part can happen when either the product of the cyclotron frequency and the transit time approaches zero or for the case of a very high magnetic field. We emphasize though the flaw in such a prediction of vanishing transverse thermopower for a high magnetic field by noting that the semi-classical approach is an inaccurate portrayal of electron motion in a strong magnetic field and masks the tacit discarding of the formation of Landau levels (LLs). We qualitatively discuss the quantum aspect of this problem and the more accurate oscillating thermopower behaviour observed with the onset of the quantum Hall regime (that owes its genesis to the formation of LLs) for high magnetic fields. From the generalized magneto-thermopower results, the magnetic field free power-factor (the product of the square of the Seebeck coefficient and electric conductivity) of PbTe thin films is easily obtainable and show a larger contribution arising from the electrons with textit{RSOC}-induced spin-down polarization. Finally, we indicate possible ways to modulate thermopower including changes to the dimensional confinement and application of strain - effects that are mirrored in the effective mass of the electrons.

\section*{Results}
The thermopower calculations begin with the pair of basic equations that connect heat and charge current. In its most general form, we can write~\cite{abrikosov2017fundamentals}
\begin{equation}
\begin{aligned}
F_{i} = \sum_{k}\rho_{ik}j_{k} + \sum_{k}Q_{ik}\dfrac{\partial T}{\partial x_{k}}, \\
q_{i} = \sum_{k}\Pi_{ik}j_{k} - \sum_{k}\chi_{ik}\dfrac{\partial T}{\partial x_{k}}.
\label{thermeq}
\end{aligned}
\end{equation}
In Eq.~\ref{thermeq}, we identify the electric field as $ \mathbf{F} $, the heat flux is denoted by $ \mathbf{q} $, the electric resistivity is $ \rho $ while $ Q\left(\Pi\right) $ is the Seebeck/thermopower (Peltier) coefficient. The last tensor, $ \chi $, denotes the thermal conductivity. These tensor quantities are simple constants for an isotropic model. However, an isotropic system can also acquire a tensorial character in presence of $ H $, a weak magnetic field. For an isotropic system, the pair of equations (Eq.~\ref{thermeq}) is modified to
\begin{equation}
\begin{aligned}
\mathbf{F} &= \rho\mathbf{j} + \sum_{k}Q_{ik}\dfrac{\partial T}{\partial x_{k}} + \nu_{1}\left(\mathbf{H} \times \mathbf{j}\right) + \nu_{2}\left(\mathbf{H} \times \nabla T\right), \\
\mathbf{q} &= \Pi\mathbf{j} - \chi\dfrac{\partial T}{\partial x_{k}} + \nu_{1}^{'}\left(\mathbf{H} \times \mathbf{j}\right) + \nu_{2}^{'}\left(\mathbf{H} \times \nabla T\right).
\label{thermB}
\end{aligned}
\end{equation}
To explain the additional terms $ \mathbf{H} \times \mathbf{j} $ and $ \mathbf{H} \times \nabla T $, note that $ \mathbf{H} $ is a pseudo-vector while $ \mathbf{E} $, $ \nabla T $, and $ \mathbf{q} $ are polar vectors; the change to the heat and electric current, therefore, must be of the form given in Eq.~\ref{thermB} to be in conformity with the general relation that the vector product of a pseudo-vector and polar vector must be a polar vector. 

\begin{figure}[!htb]
\centering
\includegraphics[scale=0.8]{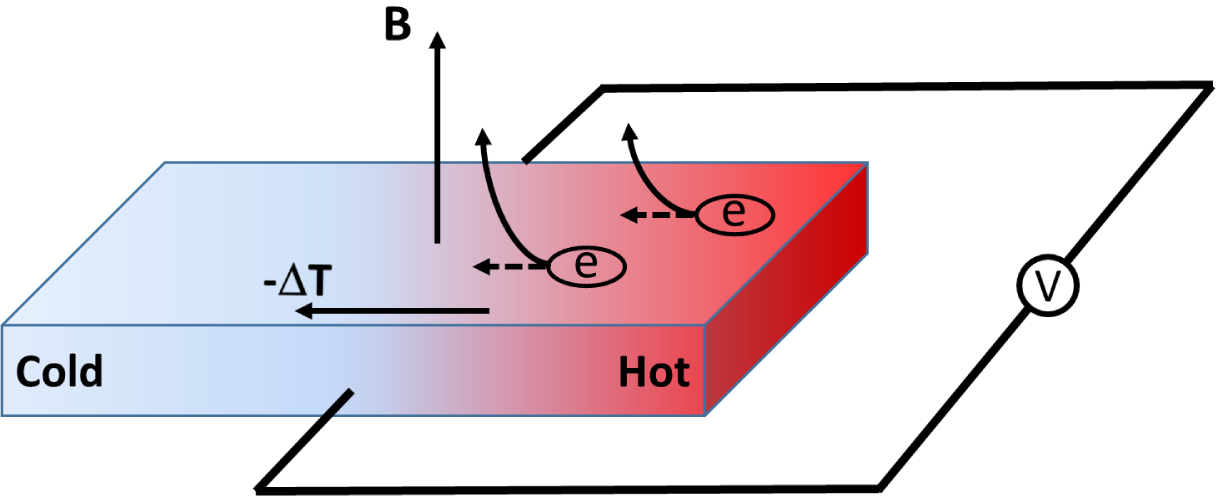}
\caption{The Nernst effect which describes an electromotive force (\textit{emf}) transverse to a temperature gradient in presence of a magnetic field (directed out-of-plane in this figure) is schematically shown here. The transverse \textit{emf} as we show later is adjustable via electron dopant density and the transport scattering time.}
\label{nsch}
\vspace{-0.3cm}
\end{figure}

The presence of a magnetic field, in addition, to the usual longitudinal thermopower can also set up an electromotive force (emf) or a thermal gradient. Two familiar examples in this regard are the \textit{Ettingshausen} and \textit{Nernst} effects; while the former describes the generation of a transverse temperature gradient (for a magnetic field directed out-of-plane and an in-plane current), the second phenomenon (Fig.~\ref{nsch}) produces an emf perpendicular to the direction of a temperature difference. As a quantitative illustration of the two aforementioned thermoelectric effects, we begin by writing out the heat equation (the second in the pair of equations in Eq.~\ref{thermB}) in component form and setting $ \partial T/\partial x = 0 $; further, assuming no heat and electric current flows along the \textit{y}-axis $\left(q_{y} = 0, j_{y} = 0\right)$, we have $ \partial T/\partial y = \left(\nu_{1}^{'}/\chi\right)Hj_{x} $. This is the Ettingshausen effect: A temperature gradient transverse to the flow of current exists in presence of a magnetic field (assumed here directed along the out-of-plane \textit{z}-axis). The Ettingshausen coefficient is $ \nu_{1}^{'}/\chi $. Likewise, for a finite temperature gradient, say, along the \textit{x}-axis $\left(\partial T/\partial x \neq 0; \partial T/\partial y = 0 \right)$ and vanishing charge currents $\left(j_{x},j_{y}\right)$, a potential drop $ \left(E_{y} = \nu_{2}H\partial T/\partial x\right) $. develops. This is the Nernst effect and the corresponding coefficient is $ \nu_{2}H $. 
A few other notable examples of the influence of a magnetic field on thermoelectric behaviour are the Righi-Leduc effect and the adiabatic Nernst and Hall phenomena~\cite{behnia2015fundamentals}. The quantitative determination of longitudinal and transverse thermopower $\left(S\right)$ can be most easily carried out within a linearized Boltzmann formalism; for non-interacting electrons, the expression (also known as the Cutler-Mott formula) is~\cite{hwang2009theory}
\begin{equation}
Q = -\dfrac{1}{eT}\dfrac{\int dE \tau\left(E\right)\left(E - \mu\right)\left(-\partial f/\partial E\right)}{\int dE \tau\left(E\right)\left(-\partial f/\partial E\right)}.
\label{seebeck}
\end{equation}
The integrals in Eq.~\ref{seebeck} can be simplified using the Sommerfeld expansion to yield a more compact form: 
\begin{equation}
Q = -\left(\pi^{2}k_{B}^{2}/3e\right)T\left(1/\sigma\right)\partial\,\sigma\left(E\right)/\partial\,E.
\label{seebcf}
\end{equation}
The derivative in Eq.~\ref{seebcf} is evaluated at the Fermi energy. To use this compact form in presence of a magnetic field which introduces anisotropy, we must change to a tensor notation. The longitudinal and transverse thermopower expressions are therefore:
\begin{subequations}
\begin{equation}
Q_{xx} = -\left(\pi^{2}k_{B}^{2}/3e\right)T\left(\rho_{xx}\sigma_{xx}^{'} + \rho_{xy}\sigma_{yx}^{'}\right).
\label{longthp}
\end{equation} 
Here, $ \rho = 1/\sigma $, and $ \sigma $ is the magneto-electric conductivity tensor. The differentiation is indicated by the primed (superscript) notation. Likewise, for the transverse part, we have
\begin{equation}
Q_{xy} = -\left(\pi^{2}k_{B}^{2}/3e\right)T\left(\rho_{xx}\sigma_{xy}^{'} + \rho_{xy}\sigma_{yy}^{'}\right).
\label{transthp}
\end{equation} 
\end{subequations}
In light of the above thermopower expressions, it is evident that we need to determine the magneto-electric conductivity which can be derived rigorously within the Boltzmann approximation; for a proof, see Chap. 8 in Ref.~\cite{nag2012electron}. However, a single band model with parabolic energy bands should essentially offer an identical result. We simply quote it (and its reciprocal, the resistivity, $ \rho $) here~\cite{grosso2014solid}
\begin{equation}
\begin{aligned}
\sigma\left(B\right) = \dfrac{ne^{2}\tau}{m^{*}}\dfrac{1}{1 + \omega_{c}^{2}\tau^{2}}\begin{pmatrix}
1 & -\omega_{c}\tau \\
\omega_{c}\tau & 1
\end{pmatrix};&
&\rho\left(B\right) = \dfrac{m^{*}}{ne^{2}\tau}\begin{pmatrix}
1 & \omega_{c}\tau \\
-\omega_{c}\tau & 1
\end{pmatrix}.
\label{mgec}
\end{aligned}
\end{equation}
In Eq.~\ref{mgec}, the relaxation time is $ \tau $, the effective mass is indicated by $ m^{*} $, and $ \omega_{c} = eB/m^{*} $ denotes the cyclotron frequency. The electron density is $ n $ in appropriate unit. The external magnetic field is included through the vector $ \mathbf{B} = \mu_{0}\mathbf{H} $, where $ \mu_{0} $ is the magnetic permeability. Inserting the magneto-conductivity tensor expression in Eqs.~\ref{longthp} and ~\ref{transthp}, the two desired thermopower quantities $ \left(Q_{xx}, Q_{xy}\right) $ can be computed. For an estimation of the relaxation time, we employ a semi-empirical approach that applies to non-degenerate semiconductors; this method varies the relaxation time as $ \tau = \lambda\varepsilon^{s} $. Here, $ \varepsilon $ is the energy while $ \lambda $ and the exponent $ s $ are scattering potential governed material constants. For cases where this semi-empirical estimation of the relaxation time is reasonably accurate, the thermopower expressions can be readily obtained. Substituting for $ \tau $ as $ \lambda\varepsilon^{s} $ and evaluating the matrix product $ \left[\rho\right]\left[\sigma^{'}\right] $ whose elements appear in Eqs.~\ref{longthp}, ~\ref{transthp}, we have
\begin{equation}
\begin{aligned}
Q_{xx} &= -\left(\dfrac{\pi^{2}k_{B}^{2}T}{3e}\right)\left(\dfrac{1}{\varepsilon_{f}}\right)\left(1 + \dfrac{s}{1 + \omega_{c}^{2}\tau^{2}}\right), \\
Q_{yx} &= -\left(\dfrac{\pi^{2}k_{B}^{2}T}{3e}\right)\left(\dfrac{1}{\varepsilon_{f}}\right)\left(\dfrac{s\omega_{c}\tau}{1 + \omega_{c}^{2}\tau^{2}}\right).
\label{sxxxy}
\end{aligned}
\end{equation} 
In writing Eq.~\ref{sxxxy} that involves the derivative of the magneto-conductivity tensor, we used the relation $ \dfrac{dn}{d\varepsilon} = g\left(\varepsilon\right) $. Here, $ g\left(\varepsilon\right) $ is the two-dimensional density of states (see Eq.~\ref{dosapp}, Methods Section). The magneto-thermopower tensor components in Eq.~\ref{sxxxy} are defined at $ \varepsilon_{f} $, the Fermi level. In passing, observe that while the thermopower components seem independent of electron density, they, however, are modulated via the $ n $-dependent $ \varepsilon_{f} $, which is Eq.~\ref{fermif1} derived in the Methods section. Further, it is easy to show using an analogous form of Eq.~\ref{transthp} that $ Q_{xy} = -Q_{yx} $.

Before we present numerical estimates for thermopower, it is instructive to enunciate on few key aspects of the expressions (for the low B-field regime, where the classical method employed remains valid) derived in Eq.~\ref{sxxxy}. First of all, note that the presence of a magnetic field (directed out-of-plane, along \textit{z}-axis) induces a transverse heat gradient and the thermopower $\left(Q\right)$, has a finite \textit{off}-diagonal component, $ Q_{xy} $. Further, for impurity scattering events that typically do not show a strong functional dependence on energy, a possible scenario when electrons are located close to the conduction band minimum, the exponent $ s $ can be approximated to zero. A proof of this appears in the Methods section. Setting $ s = 0 $ in Eq.~\ref{sxxxy}, we recover thermopower expressions under a vanishing magnetic field~\cite{obraztsov1964thermomagnetic,fletcher1999magnetothermoelectric}. For the other extreme case of high-field limit, such that $ \omega_{c}\tau\,\rightarrow \infty $, it is easy to verify that $ Q_{xy} $ ceases to exist, much like when the magnetic field is turned off. However, in this case, we must exercise caution (also see the discussion presented in the summary) that the putative expressions in Eq.~\ref{sxxxy} may be incorrect inasmuch as their formulation goes at high magnetic fields. The quantization of the electron orbit via formation of quantum mechanical Landau levels beginning at high magnetic fields and the possibility of the onset of quantum hall features radically alters the physical description and therefore any analysis (similar to the one presented) in this regime that has its genesis in classical arguments would be mostly erroneous. Further, note that expressions for the thermopower tensor components in a magnetic field can be reasonably well-estimated if accurate material-specific values such as the effective mass and relaxation time $\left(\tau\right)$ are at hand. To elucidate, we considered the relaxation time tailored to a variety of scattering events by adjusting the exponent, $ s $. As an illustration of the process that fixes the value of $ s $, we illustrate the specific case of non-interacting impurities and compute the relaxation time using a self-consistent Born approximation (SCBA). The SCBA (see Eq.~\ref{scba1} in the Methods section and the accompanying discussion that follows) furnishes the relaxation time from the imaginary part of the self-energy of electrons whose motion is disturbed by the impurity scattering potential~\cite{rammer2004quantum}. A calculation of the relaxation time via SCBA and other required parameters must proceed by setting up an appropriate Hamiltonian for the thermoelectric material accounting for dimensional-effects and composition of the bulk/nanostructure. The numerical diagonalization of the appropriate Hamiltonian supplies the dispersion relation useful in the evaluation of the effective mass, assigning the relevant energy scale of the problem, and an assessment of the part played by scattering that alters carrier mobility. The Methods section outlines the Hamiltonian (Eq.~\ref{kp1}) for quantum wells of lead salts, the target thermoelectric material. 

As for the semi-empirical approach, it is evident from Eq.~\ref{sxxxy} that an enhanced thermopower is achievable if the exponent $\left(s\right)$ acquires a higher number; in fact, suggestions~\cite{goldsmid1960applications} have been made to dope PbTe, a well-studied thermoelectric such that $ s $ is reinforced via impurity-scattering. This scheme, however, works insofar as there is no obvious degradation of electron mobility, for the efficient generation of thermoelectric currents finds expression in figure-of-merit $\left(ZT\right)$ which relies on conductivity and can suffer a considerable reduction in an attempt to augment the scattering. However, it is remarkable that a boost to scattering (and consequently a higher $ s $) led to an observable increase in $ ZT $ for bismuth structures~\cite{issi1976electron}. In this context, also note how for scattering events which do not have an energy dependence $\left(s = 0\right)$ or simply a constant relaxation time, the transverse magneto-thermopower or the so-called \textit{Nernst-Ettingshausen} effect vanishes. In practice though, scattering events are always energy-dependent phenomena; however, as an approximation, Bloch electrons close to the conduction band minimum which are scattered by constant spherical delta potentials of impurities, the relaxation time can be shown to be a fixed number. The relevant relaxation time calculation which proves this constancy is presented in the Methods section (Eq.~\ref{scba1} and the following analytic results).

\begin{figure}
\centering
\includegraphics[scale=0.8]{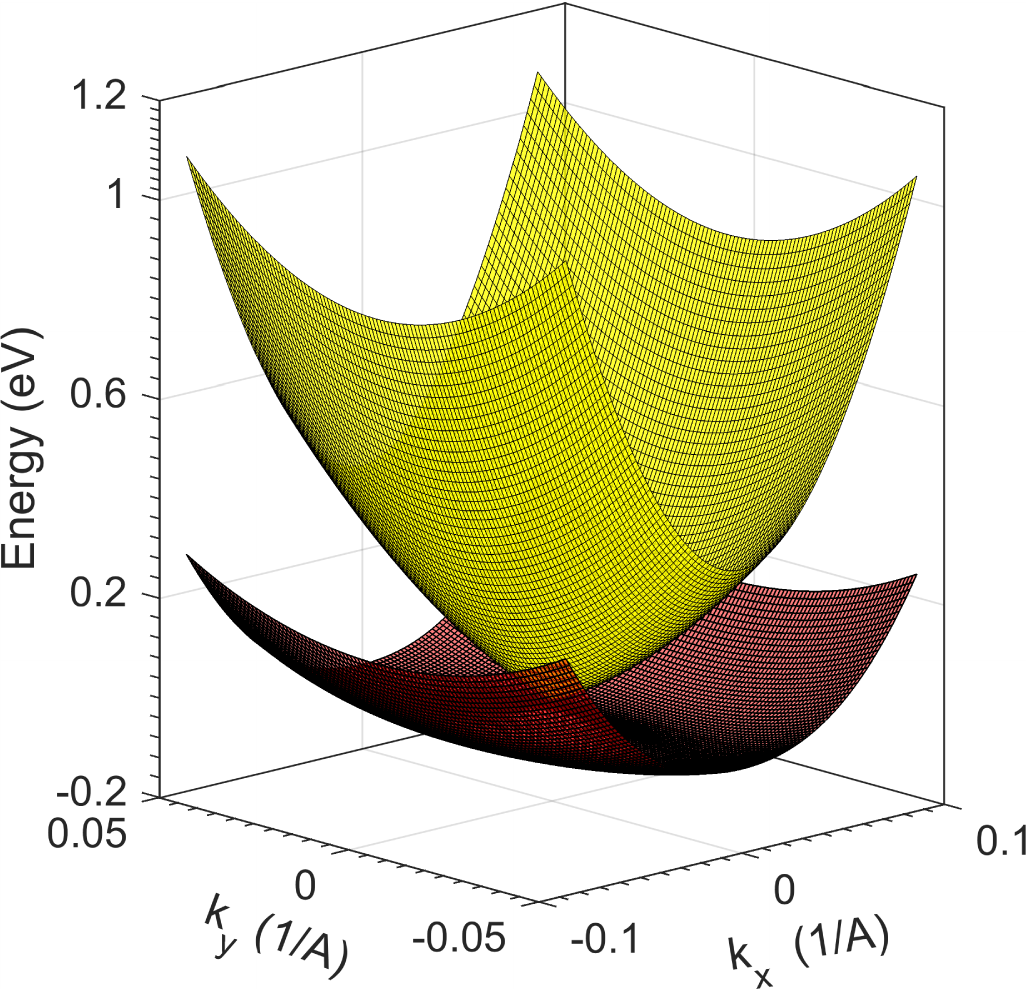}
\caption{The Rashba spin split Fermi surfaces of the conduction electrons located close to the \textit{L}-valley of a $ 6.0\, nm $ wide PbTe quantum well. PbTe, fulfilling the requirements of a narrow band gap and large intrinsic spin-orbit coupling, serves as an ideal representative material to illustrate extrinsic Rashba splitting. The spin-polarized conduction electron ensembles are modeled by the simple dispersion relation $ \varepsilon = \hbar^{2}k^{2}/2m^{*} \pm \lambda_{R}k $. The energetically higher band profile depicts the spin-up carriers. Note that for greater visual clarity, the Rashba coefficient was artificially increased to $ \lambda = 2.0\, eV\AA $ and $ \vert k \vert = 0.1\, \AA $. An accurate estimate of the Rashba coefficient and calculation of the effective mass of the PbTe quantum well conduction electrons $\left(0.0565m_{0}\right)$ are described in greater detail in the Methods section.}
\label{rashsp}
\end{figure}

Finally, observe from Eq.~\ref{sxxxy} the possibility of carrier ensembles displaying a thermopower response to temperature and potential gradients tied to their respective Fermi levels. A common occurrence of distinguishable Fermi levels within a material can be readily recognized for spin-split Bloch electrons, where the splitting emerges from spin-orbit coupling (\textit{soc} that manifests either from an intrinsic atomic contribution or the externally induced Rashba-effect (\textit{RSOC}). In general, the intrinsic \textit{soc} is significant in heavy metals with a large nucleus while quantum wells and thin films of narrow gap materials show a pronounced \textit{RSOC} splitting the Fermi surface into two groups of spin-polarized electrons. A simple realization of this can be seen (see Fig.~\ref{rashsp}) by considering the minimal quadratic Hamiltonian with the linear Rashba contribution~\cite{peres2011spin}
\begin{equation}
H_{rs} = \dfrac{p^{2}}{2m^{*}} + \lambda_{R}\left(\sigma_{x}k_{y} - \sigma_{y}k_{x}\right),
\label{cbrs}
\end{equation}
where $ \lambda_{R} > 0 $ is the Rashba coefficient and determines the robustness of the splitting. The related dispersion is of the form $ \varepsilon = \hbar^{2}k^{2}/2m^{*} \pm \lambda_{R}k $; the splitting is therefore $ 2\lambda_{R}k $, where $ k $ is the in-plane wave vector given by $ \sqrt{k_{x}^{2} + k_{y}^{2}} $. The Fermi surfaces for the spin-up and spin-down carriers are therefore non-degenerate, the difference in their energies determined by $ \lambda_{R} $. A straightforward calculation connects the dependence of the surface Fermi energy $\left(\epsilon_{f}\right)$ to \textit{RSOC} via $ \lambda_{R} $ for a certain electron density $\left(n\right) $. We quote the result here (see Methods section for a proof) 
\begin{equation}
\epsilon_{f} = \left[2\pi\alpha\left(\sqrt{\dfrac{\lambda_{R}^{2}}{16\pi^{2}\alpha^{3}} + \dfrac{n}{\pi\alpha}} \pm \dfrac{\lambda_{R}}{4\pi\alpha^{3/2}}\right)\right]^{2}.   
\label{fermif}
\end{equation}
The upper (lower) sign is for spin-up (down) electrons and $ \alpha = \hbar^{2}/2m^{*} $. The Fermi energy (for a specific $ n $) is evidently governed by the strength of \textit{RSOC}. The corresponding vector (for later use) is
\begin{equation}
k_{f} = \dfrac{\sqrt{\lambda_{R}^{2} + 4\alpha\epsilon_{f\pm}} \mp \lambda_{R}}{2\alpha}.  
\label{fermivec}
\end{equation}
The upper (lower) sign in Eq.~\ref{fermivec} is for the spin-up (down) ensemble. A more elaborate set of remarks on this aspect and numerical calculations in context of realistic quantum well systems is included in the Methods section.  
We can now insert the Rashba-split Fermi energy expressions (Eq.~\ref{fermif}) in Eq.~\ref{sxxxy} and reasonably expect a spin-dependent magneto-thermopower. This could be essentially thought of as the thermal analog of the spin-dependent \textit{Seebeck} phenomenon and may likewise be referred to as the spin-dependent \textit{Nernst-Ettingshausen} effect. In what follows, we employ this framework to examine the spin-delineated magneto-thermoelectric tensor quantities and the related power factor (for a zero $ \mathbf{B} $ field case) of lead-salt quantum wells. 

\subsection*{Lead salt films}
\begin{figure}[!t]
\centering
\includegraphics[scale=1.2]{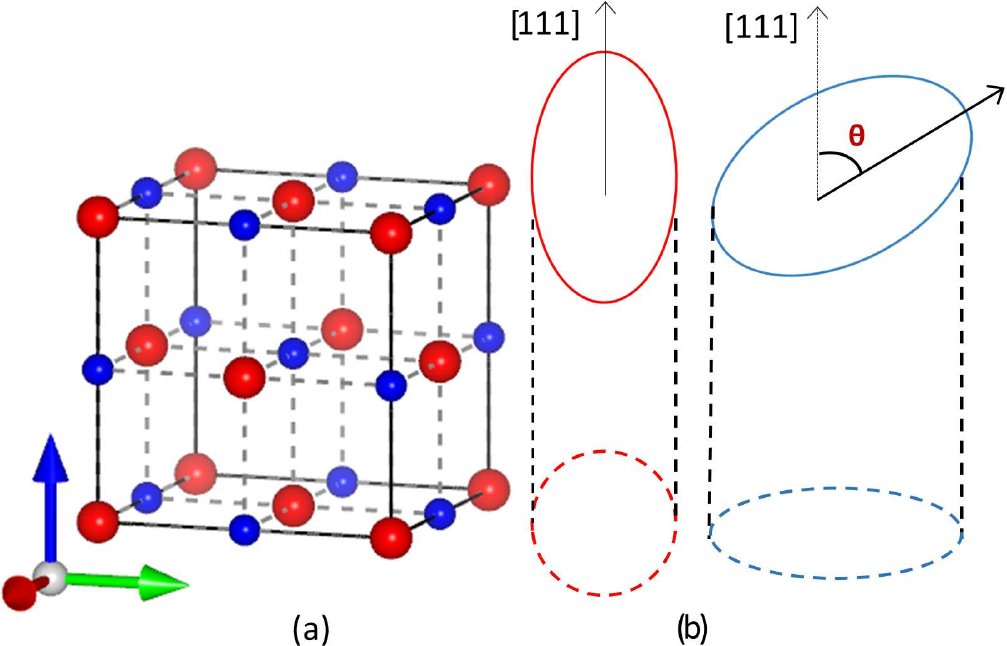}
\caption{The left panel (a) shows the representative rock salt crystal structure of PbTe, where the smaller atom is Te (blue) and Pb (red) with a higher atomic number (82) is identified by a larger size. The Te atom in this arrangement is sandwiched between two layers of Pb, which is the cation. PbS and PbSe have identical crystal structures. The dotted lines show the bonds between the Te and Pb atom. The right panel (b) depicts the momentum space in $ \left[111\right] $-grown bulk samples around the four-fold degenerate $ L $-valleys where the lead salts have a direct fundamental gap. However, three such valleys are oblique with their long axes tilted (at an angle $ \theta = 70.5^{\circ} $) to the \textit{z}-direction which is aligned along the $ \left[111\right] $ space vector. Additionally, a single valley (referred to as the `longitudinal' valley) has its axis parallel to $ \left[111\right] $, the chosen \textit{z}-direction in the current set of reference coordinates. The projection of the tilted valleys in the two-dimensional plane is an ellipse-shaped Fermi surface denoting anisotropy while the longitudinal valley projects as a perfect circle. The projections in both cases are marked by solid dotted lines. Note that for a $ \left[001\right] $-grown sample, all four valleys are tilted at an angle $ \theta = 53^{\circ} $ to the current \textit{z}-axis.}
\label{pbstr}
\vspace{-0.4cm}
\end{figure}

We alluded to the utility of tellurides above as desirable thermoelectric materials. For numerical calculations we select PbTe as the candidate telluride, the choice of which is primarily driven by its well-regarded thermoelectric performance and a narrow band gap coupled to a strong spin-orbit coupling. The latter two attributes makes it suitable to observe spin-dependent thermoelectrics. The derived spin-resolved magneto-thermopower results here though are general and can be applied to a wide variety of materials by selecting the right effective mass and the correct form of the \textit{RSOC} Hamiltonian. It is also worth noting that the lead chalcogenides, PbS and PbSe, are iso-structural with PbTe crystallizing in a rock-salt crystal arrangement and are expected to show similar thermoelectric behaviour. Briefly, the lead salts crystallize in the rock-salt structure (a two-atom basis) which consists of two inter-penetrating face-centred-cubic (fcc) lattices displaced by $ \left(0.5, 0.5, 0.5\right)\times a $, where $ a $ is the lattice constant~\cite{khokhlov2002lead}. They are marked by narrow direct band gaps at four equivalent $ L $ valleys where the conduction and valence band extrema occur. For samples grown on [111]-oriented substrates, the long-axis (of the $ L $ valley which is an ellipsoid) of one of $ L $-valleys (known as the longitudinal valley) is normal to the substrate surface (see Fig.~\ref{pbstr}) while the rest three (referred to as the oblique-valleys) are tilted by $ \theta = 70.5^{\circ} $. Quantum confinement in a film, however, removes the degeneracy at the $ L $ point of the Brillouin zone pushing the three oblique valleys to a higher energy over the solitary longitudinal counterpart. In this report, we ignore such distinctions and work with the longitudinal valley tacitly assuming that the other three equally contribute to magneto-thermopower.

\begin{figure}[!t]
\centering
\includegraphics[scale=0.8]{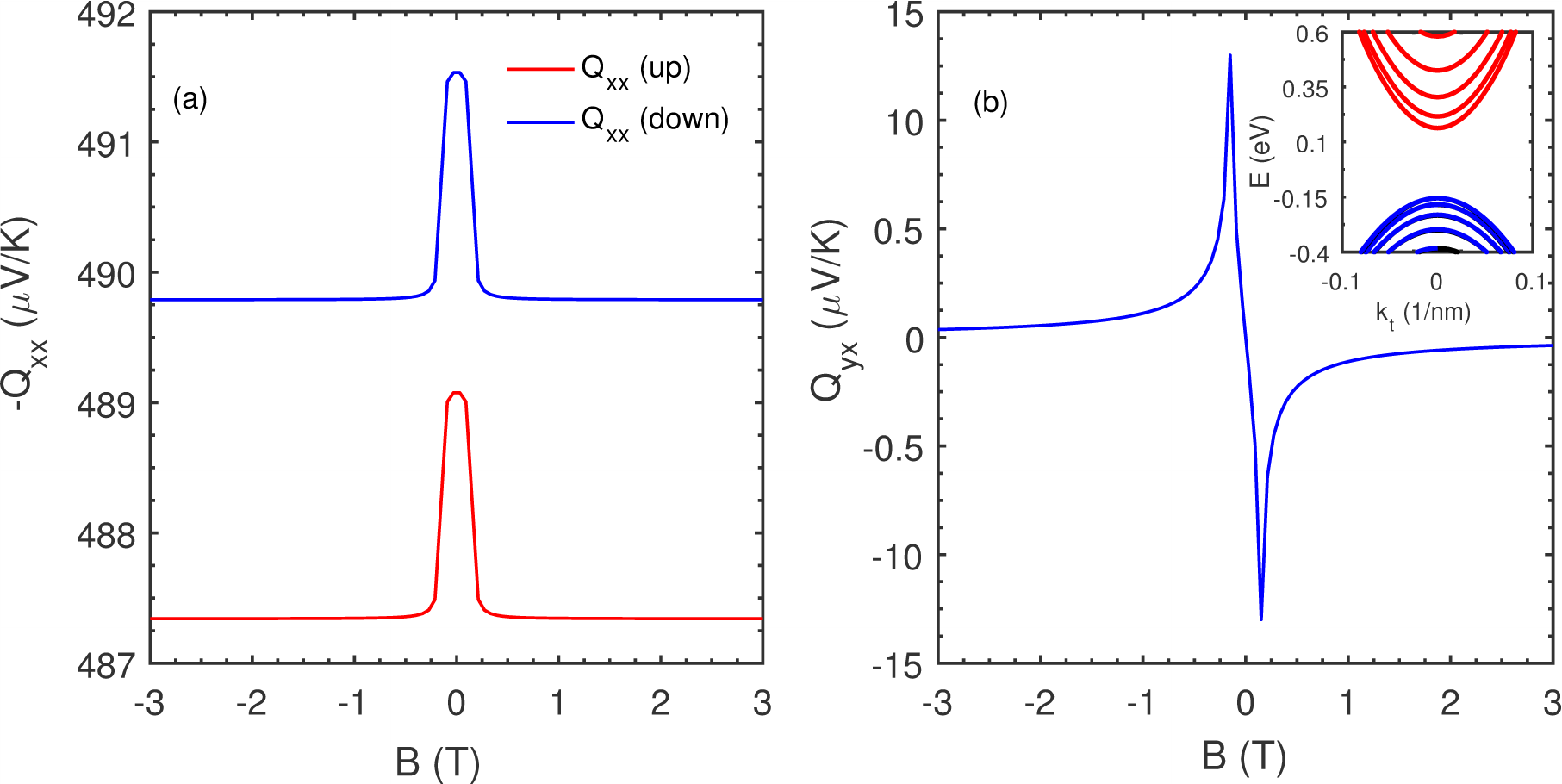}
\caption{The spin-resolved low-temperature longitudinal (a) and transverse magneto-thermopower (b) is plotted for several magnetic fields for electrons located in the \textit{L}-valley of a $ 6.0\,nm $ wide quantum well confined along the [111] axis. The contribution of the oblique valleys is not included in this plot. For a quantitative result, in accord with our use of Mott's thermopower formula valid in the low-temperature regime, $ T $ was set to 4 K. The carrier density in the well was assumed to be $ n = 10^{12}\,cm^{-2} $ and the scattering time for $ s = 0.7 $ (see text for an explanation) is $ 0.1\,ns $. Numerically, the transverse thermopower for the Rashba spin-polarized electron ensembles are nearly identical while a difference is observable in the case of its longitudinal counterpart. The inset shows the dispersion of the PbTe quantum well from which we calculate the transverse effective mass to be 0.0565m$_{0}$. The Rashba-spin splitting energy was calculated using Eq.~\ref{rasz} by a simple insertion of appropriate lead salt band parameters noted in the Methods section.}
\label{thsxxy}
\vspace{-0.3cm}
\end{figure}

For a numerical evaluation of the magneto-thermopower in thin films of lead salts, we start by fixing certain quantities. The carrier (electron) concentration~\cite{seln} is set to $ n = 10^{11}\,cm^{-2} $ and the temperature is $ T $ = 20 K for all calculations. Note from Eq.~\ref{fermif} the presence of two spin-polarized Fermi surfaces for a definite concentration. The parameters we vary are the carrier effective mass $\left(m^{*}\right)$, the magnetic field strength, and the choice of scattering mechanism that influences the relaxation time. As we qualitatively discuss later, $ m^{*} $ is calibrated by selecting different film thicknesses and alloying ($ x $ is the mole fraction) with SnTe to yield Pb$_{x}$Sn$_{1-x}$Te. The implication of using Pb$_{x}$Sn$_{1-x}$Te in addition to a ready source of composition-controlled tunable energy-levels lies in a band gap closing feature beyond a threshold tin concentration. This is discussed more fully in the closing section. For definiteness, we consider pristine and strained PbTe films and plot the spin-resolved longitudinal and transverse thermopower of conduction electrons in $ 6.0\,nm $ and $ 8.0\,nm $ thick films. The effective mass in Eq.~\ref{cbrs} for all cases are calculated using a $ 4 \times 4 $ \textit{k.p} Hamiltonian (Eqs.~\ref{hrs} and ~\ref{kp1}) reviewed in the Methods section. The external spin-splitting (\textit{RSOC}) in each case is determined assuming an asymmetry-induced out-of-plane electric field $\left(\mathbf{F}\right)$ operates; setting $ \mathbf{F} = 10^{6}\, V/m $, we obtain the Rashba coefficient (Eq.~\ref{rasz} in the Methods section) and the spin-polarized energy states. With these simulation parameters in mind, the quantities $ Q_{xx} $ and $ Q_{xy} $ are shown for several magnetic field strengths in Fig.~\ref{thsxxy}.
In preparing Fig.~\ref{thsxxy} for the magneto-thermopower contribution of the conduction electrons of thin films of lead salts located close to the longitudinal $ L $-valley minimum, we used a $ 6.0\,nm $ PbTe quantum well (QW). The band parameters for this thin film structure including the transverse effective mass of the conduction electrons and the direct band gap were determined to be 0.0565*m$_{0}$ (the free electron mass is m$_{0} $) and $ 0.2131\, eV $. 

We now comment on the nature of the plots and choice of other variables used: First of all, observe that the longitudinal thermopower does not change sign as the magnetic field is reversed while the transverse component is odd in $ \mathbf{B} $, a result which readily follows from the nature of the conductivity expressions in Eq.~\ref{mgec}. The longitudinal conductivity and resistivity are even while their transverse counterparts are odd functions of $ \mathbf{B} $ bestowing an even and odd character to each term in $ Q_{xx} $ and $ Q_{yy} $, respectively. The relaxation time in magneto-thermopower expressions was set to $ 0.1\, ns $ and the scattering exponent $ s = 0.7 $. In this very approximate scheme of relaxation time determination, we are heuristically guided by standard empirical formulations that predict scattering exponents of $ s = 0.5 $ and $ s = 2 $ for neutral and ionized impurities, respectively~\cite{nag2012electron}. Scattering off acoustic phonons are modeled by letting $ s = -0.5 $. Since most phonon modes would be suppressed at low temperatures, we choose an intermediate value of $ s = 0.7 $ which lies between the assigned exponents $\left(0.5 < s < 2\right)$ for neutral and charged impurity scattering. Lastly, observe that for the operational Rashba spin splitting, the transverse thermopower is closely matched for both spin ensembles in contrast to the longitudinal case where the spin-up conduction electrons exhibit a higher value. This result is in agreement with the expression derived in Eq.~\ref{sxxxy} that shows an inverse relationship to carrier energy translating into a higher longitudinal thermopower for spin-up electrons. The spin-up electrons occupy an energetically higher Fermi surface in comparison to their oppositely polarized counterpart. In context of Fig.~\ref{thsxxy}, the QW conduction electron concentration was assumed to be $ n = 10^{12}\, cm^{-2} $ from which the \textit{RSOC} splitting energy $\left(2\lambda_{R}k\right)$ is $ 12.0\, meV $ for $ \mid k \mid = 1.0\, \AA $. In the context of \textit{RSOC}-induced splitting, we briefly digress here to emphasize on a key PbTe material parameter. Firstly, note that the Rashba parameter $\left(\lambda_{R}\right)$ is given by the relation $ \lambda_{R} = \lambda_{0}E_{z} $, where $ \lambda_{0} $ is a material-specific number (see Eq.~\ref{rasz}, Methods section) and $ E_{z} $ is the magnitude of a structural inversion asymmetry (SIA) generated~\cite{ganichev2004experimental} out-of-plane electric field. For estimating $ E_{z} $ supposing that it arises only from the charge density in the QW, it is given as $ \mathbf{E} = en_{s}/\epsilon $, where $ n_{s} $ is the charge density and $ \epsilon $ is the static dielectric constant; for PbTe, the static dielectric constant is $\approx $ 420. The implications of this large dielectric constant are profound and briefly touched upon in the following sub-section. Before closing, to be clear again, the spin-dependent thermopower expressions are a single-particle phenomenon where two channels (for spin-up and spin-down) independently carry heat and spin, unlike, say, for example, the spin-Seebeck effect which is believed to be a magnon-driven spin current affair~\cite{xiao2010theory,bauer2012spin}.

\subsection*{Power factor of Pb-based thin films}
An important marker in gauging thermoelectric processes is the power factor defined (for zero magnetic fields) as the product of $ Q^{2} $ and electric conductivity. The longitudinal and transverse Seebeck coefficients (for the zero magnetic field case) and can be easily obtained by setting the cyclotron frequency $\left(\omega_{c}\right)$ to zero in Eq.~\ref{sxxxy}. Straightforwardly, the transverse thermopower vanishes while the longitudinal component $\left(Q_{xx}\right)$ reduces to:
\begin{equation}
Q_{xx} = -\left(\dfrac{\pi^{2}k_{B}^{2}T}{3e}\right)\left(\dfrac{1}{\varepsilon_{f}}\right)\left(1 + s\right).
\label{nobth}
\end{equation}
The utility of the Seebeck coefficient lies in the formulation of the figure of merit generally expressed as $ ZT = \dfrac{Q^{2}\sigma}{\kappa}T $, where $ \sigma $ is the electric conductivity and $ \kappa $ is the combined lattice and electronic contribution to thermal conductivity. Evidently, for a higher $ ZT $, which is desirable for improved thermoelectrics, the product $ Q^{2}\sigma $ must be maximized while ensuring a low thermal conductivity. A concurrent fulfillment of this dual set of conditions is purportedly a difficult proposition, however, lead salts offer much promise and multiple pathways in achievement of this goal~\cite{harman1996high,heremans2017tetradymites}. To elucidate on this point, first notice that electric conductivity $ \sigma $ is a charge density dependent quantity $ \left(\sigma = ne\mu\right) $ which in case of Pb-salts is augmented by the valley degeneracy $\left(g = 4\right)$ of the $ L $ high-symmetry point in the Brillouin zone. The conductivity also receives more augmentation from the mobility $\left(\mu\right)$ which is large for low effective masses. The low effective masses are an outcome of the strong intrinsic spin-orbit coupling. In alliance with the favourable microscopic arrangement, the highly polarizable bands in lead salts give rise to a large static dielectric constant~\cite{alves2013lattice} leading to a significant Bohr radius $\left(a_{B}\right)$. A large $ a_{B} $ serves to conceal impurities and imperfections that may lower the overall mobility and therefore supports a high electric conductivity. It is then reasonable to expect a substantial power factor for thin films of lead salts. In the following, we describe a set of steps to estimate it; as before, PbTe is the representative material.

\begin{figure}
\centering
\includegraphics[scale=0.85]{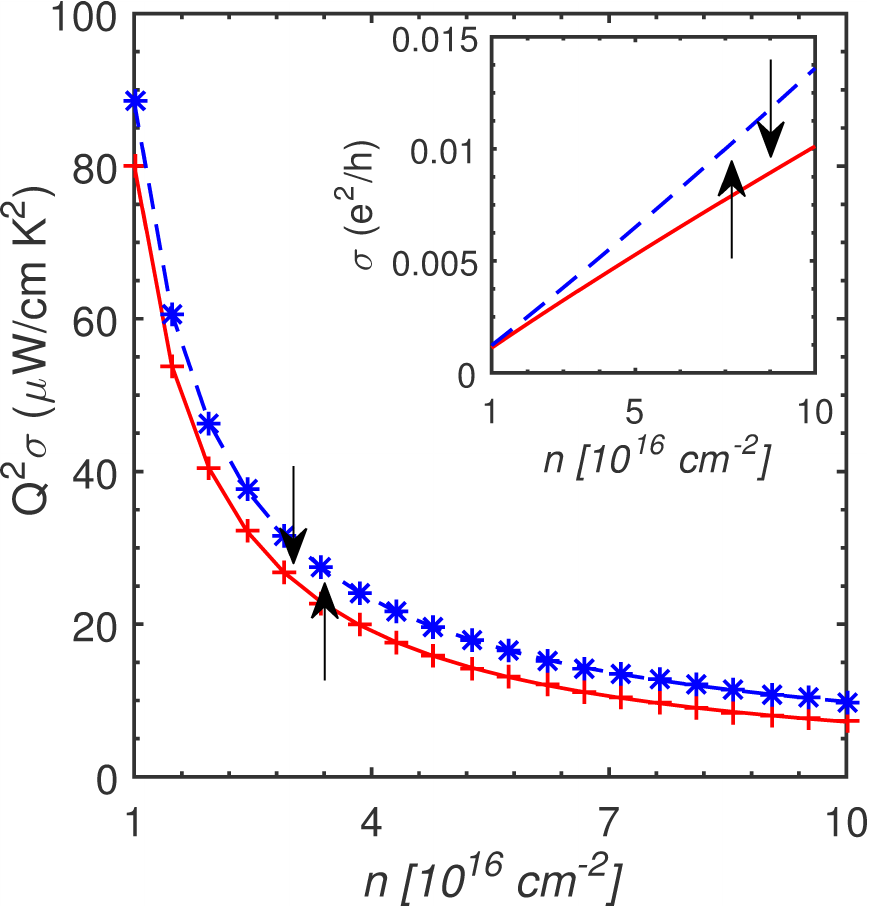}
\caption{The spin-resolved power factor (\textit{pf}) of a $ 6.0\, nm $ wide PbTe QW is plotted for $ T = 1\, K $ as a function of carrier density. The \textit{pf} is higher for the spin-down ensemble. This is also reflected in the enhanced charge conductivity (inset) of the spin-down branch calculated using Eq.~\ref{sigbz}. No electrostatic screening was included in the numerical estimation of the power factor and conductivity. The impurity scattering time $\left(\tau\right)$ was adjusted to $ 1.0\, ns $.}
\label{pfpic}
\end{figure}

A semi-classical approximation of the conductivity, a linearized Boltzmann equation within the relaxation time approach allows the conductivity to be written as
\begin{equation}
\sigma = \dfrac{e^{2}v_{f}^{2}}{2}\int\, d\varepsilon D\left(\varepsilon\right)\tau\left(\varepsilon\right)\left(-\dfrac{\partial f}{\partial \varepsilon}\right).
\label{sigbz}
\end{equation}
The Fermi velocity is $ v_{f} $, density of states (DOS) is $ D\left(\varepsilon\right) $, and $ \tau $ is the relaxation time in Eq.~\ref{sigbz}. For low-temperatures, replacing the Fermi distribution with the step-function simplifies it to $ \sigma = e^{2}v_{f}^{2}D\left(\varepsilon\right)\tau\left(\varepsilon\right)/2 $. The Fermi velocity is $ v_{f} = \hbar k_{f}/m^{*} $, where $ k_{f} $ is given by Eq.~\ref{fermivec}. Note that in expressing the velocity operator, we have ignored the Rashba contribution and only retained the parabolic part of the Hamiltonian. The relaxation time can be gauged by considering impurity scattering and excluding any phonon-assisted disruption. The relaxation time for such a condition can be written as $ 1/\tau = \left(2/\hbar\right)n_{i}m^{*}v_{i}^{2}/2\hbar^{2} $. The quantity $ n_{i} $ is the impurity density and $ v_{i} $ the corresponding strength. Putting Eq.~\ref{nobth}, the zero-temperature form of Eq.~\ref{sigbz}, and the expression for $ \tau $ together, the power factor $\left(Q^{2}\sigma\right)$ can be written as
\begin{equation}
\left(Q^{2}\sigma\right)_{\pm} = \dfrac{\pi^{3}k_{B}^{4}T^{2}}{18m^{*}}\dfrac{\left(1+s\right)^{2}}{\varepsilon_{f\pm}^{2}}k_{f}^{2}\tau,
\label{pffin}
\end{equation}
where the upper (lower) sign is, as usual, for the spin-up (down) ensemble. We have also approximated the DOS  discarding the Rashba contribution and subsituted for $ D\left(\varepsilon\right) $ as $ m^{*}/2\pi\hbar^2 $. For a quantitative prediction of the power factor, we use a $ 6.0\, nm $ wide PbTe QW in Fig.~\ref{pfpic}. The relaxation time $\left(\tau\right) $ is approximated by setting the impurity density to $ n_{i} = 2.2 \times 10^{9} cm^{-2} $ while the corresponding potential is assigned the value, $ v_{i} = 0.1\, eVA^{2} $. As before, the \textit{L}-valley conduction electron effective mass for this QW is $ m^{*} = 0.0565m_{0} $. The relaxation time using the relation noted above for the selected impurity parameters and effective mass is roughly $ 1.0\,ns $. The power factor for a $ 6.0\, nm $ PbTe QW is displayed in Fig.~\ref{pfpic}. 

The low-temperature power factor calculation presented here represents the optimal case; impurities and related surface scattering events tend to lower the conductivity. The experimental power factor is therefore expected to be reduced vis-\`a-vis the current theoretical estimate. In fact, the highest reported power factor in PbTe QWs $ \left(n \approx 10^{24}\,cm^{-3} \right) $ at room temperature is 130 $ \mu W/cm K^{2} $.~\cite{harman1996high} Further, notice that the effective mass appears in Eq.~\ref{pffin} pointing to the distinct possibility of reduction in QW dimension or greater confinement, for instance, in a nanowire to modulate the power factor. Indeed, such observations have been extensively reported to tune the power factor; however, for another viewpoint on the aspect of dimensional analysis of this problem that also takes into account the thermal de Broglie wave length, see Ref.~\citenum{hung2016quantum}. 

\section*{Final remarks}
We have presented a semi-classical analysis of magneto-thermopower of a two-dimensional electron gas system in a PbTe quantum well. The calculations were done in the low-temperature regime and moderate magnetic fields extending up to $ \mathbf{B} = 3.0\, T $ on either side of the zero mark. Quantum mechanical effects were incorporated by extracting a realistic effective mass for the conduction electrons of the PbTe quantum well from an appropriate \textit{k.p} Hamiltonian. However, it is important to clarify that at higher magnetic fields than those considered here, there exists a sharp departure from classical behaviour. The electrons begin to circulate in Landau levels (LL) and may further quantize to reach the integer quantum Hall (IQH) regime. We have ignored the quantum mechanical coupling of the magnetic field to electron motion and the attendant changes to longitudinal and transverse magneto-conductance, especially when the integer QH effect (IQHE) sets in. Succinctly, for a pre-defined Fermi level positioned between two consecutive LLs - at a high magnetic field and low-temperature - the transverse conductivity is quantized as $ \sigma_{xy} = \nu e^{2}/h $, where $ \nu $ is the filling factor while the longitudinal tensor component ceases to exist $\left( \sigma_{xx} = 0 \right) $. By sweeping the Fermi level between energetically higher LLs, the well-known stair-like behaviour indicating a higher $ \nu $ is observed. For regions where the Fermi level is placed between two successive LLs, the constant conductivity, by a simple application of Eq.~\ref{sxxxy} leads to a vanishing transverse magneto-thermopower. Simultaneously, the longitudinal part $\left(Q_{xx}\right)$ is zero by virtue of $ \sigma_{xx} = 0 $ in the IQH regime. In the other scenario, when the Fermi level is adjusted to align with a certain LL, both the transverse and longitudinal thermopower are restored as neither $ \sigma_{xx} $ vanishes nor does $ \sigma_{xy} $ displays quantization effects. Briefly, the successive vanishing and renewed establishment of thermopower as the Fermi level is progressively adjusted between successive LLs give an overall oscillatory pattern. A plot elucidating this behaviour for 2DEG in GaAs in the IQH regime was presented by Jonson and Girvin in Ref.~\cite{jonson1984thermoelectric}.

In closing, we have theoretically analyzed the magneto-thermopower of lead salts that crystallize in the rocksalt structure. Tellurides of Pb-salts in various forms have been recognized as thermoelectrics with high efficiency, which can be further enhanced through alloying, dimensional confinement, tuning the carrier effective mass, and their large dielectric constant. Beyond these physical quantities and design parameters, the narrow band gap and strong intrinsic spin-orbit coupling contribute to a strong Rashba spin-splitting, which within the scope of the presented model leads to quantitatively different thermopower and power factor setups. It is also fitting to note here that quantum wells of Pb-salts, in particular, PbTe alloyed with SnTe show topological insulator behaviour; a quantum state of matter that has been theoretically predicted (and observed in few experimental demonstrations) to offer a much higher $ ZT $ than hitherto possible with conventional materials. As a final observation, these calculations can be transferred (with a modified minimal Hamiltonian of the type used in Eq.~\ref{cbrs}) to another well-known class of thermoelectrics - the rhombohedrally crystallizing binary tetradymites - that carry similar properties as the lead salts including a large spin-orbit coupling driven topologically non-trivial surface bands, low band gap, and high dielectric constant. \textcolor{red}{A notable tetradymite is Bi$_{2}$Te$_{3}$ that complies with all the aforementioned PbTe attributes, however, the presence of topological surface states in thin films complicates the measurement and interpretation of experimental data. The Bi$_{2}$Te$_{3}$ structures are also beset by a large concentration of impurities that enter the growth process modifying intrinsic character.}

\section*{Methods}

\subsection*{Dispersion calculations}

A minimal Hamiltonian that captures the basic band dispersion of the conduction electrons in a lead chalcogenide quantum well is 
\begin{equation}
H_{RS} = \dfrac{p^{2}}{2m^{*}} + \lambda_{R}\left(\sigma_{x}k_{y} - \sigma_{y}k_{x}\right),
\label{hrs}
\end{equation}
where we have included the linear Rashba spin-orbit interaction. The Rashba coupling constant is $ \lambda_{R} > 0 $ and is particularly strong in narrow band gap materials with strong intrinsic spin-orbit coupling, such as the lead chalcogenides. In particular, the strength of the Rashba coupling coefficient is $ \lambda_{R} = \lambda_{0}\langle\,E\left(z\right)\rangle $, where $ \langle\,E\left(z\right)\rangle $ serves as the average electric field. The material-dependent $ \lambda_{0} $ is given as~\cite{e1994spin}
\begin{equation}
\lambda_{0} = \dfrac{\hbar^{2}}{2m^{*}}\dfrac{\Delta}{E_{g}}\dfrac{2E_{g}+\Delta}{\left(E_{g} + \Delta\right)\left(3E_{g} + 2\Delta\right)}.
\label{rasz}
\end{equation}
\begin{figure}[t!]
\centering
\includegraphics[scale=1.25]{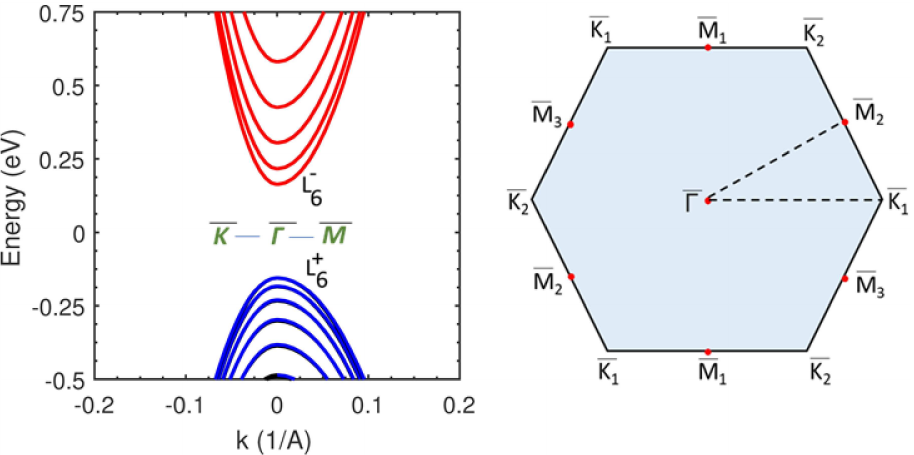}
\caption{The \textit{k.p} calculated dispersion along the $ \overline{K} - \overline{L} - \overline{\Gamma} $ path for a $ 6.0\,nm $ wide PbTe quantum well is shown on the left panel. The \textit{k.p} Hamiltonian employed describes the longitudinal valley whose axis lies along the $ \left[111\right] $ vector. The \textit{L}-valley conduction band minimum and valence band maximum $\left(L_{6}^{+}\right)$ have opposite parities described by the $ L_{6}^{-} $ and $ L_{6}^{+} $ symmetry notation. The right panel is a schematic representation of the hexagonal two-dimensional $\left[111\right]$ surface Brillouin zone of rock salt PbTe. The lettered notation at the hexagon corners denotes the high-symmetry points and the dotted triangle is the chosen path of the plotted the dispersion in the left panel.}
\label{dsbz}
\end{figure}
In Eq.~\ref{rasz}, the band gap at $ \Gamma $ is $ E_{g} $, the spin-orbit splitting is $ \Delta $ and $ m^{*} $ is the effective mass at points in momentum space. Note that the Dresselhaus coupling term is absent since bulk inversion symmetry is preserved for crystals which have rock salt crystal arrangement. The effective mass of band carriers in a lead chalcogenide quantum well can be derived from a continuum bulk $ 4 \times 4 $ \textit{k.p} model valid in the vicinity of the high-symmetry \textit{L}-valley. In the spin-resolved $\left(\uparrow\,\downarrow\right)$ basis set of $ L_{6}^{-}\uparrow $, $ L_{6}^{-}\downarrow $, $ L_{6}^{+}\uparrow $, and $ L_{6}^{+}\downarrow $, where $ L_{6}^{\pm} $ are the Bloch functions of the conduction (-) and valence band (+) edge, it takes the form~\cite{kang1997electronic} 
\begin{flalign}
H = \begin{pmatrix}
\dfrac{E_{g}}{2} + \dfrac{\hbar^{2}k_{t}^{2}}{2m_{t}^{-}} + \dfrac{\hbar^{2}k_{z}^{2}}{2m_{l}^{-}} & 0 & \dfrac{\hbar}{m}P_{l}k_{z} & \dfrac{\hbar}{m}P_{t}k_{-} \\
0 & \dfrac{E_{g}}{2} + \dfrac{\hbar^{2}k_{t}^{2}}{2m_{t}^{-}} + \dfrac{\hbar^{2}k_{z}^{2}}{2m_{l}^{-}} & \dfrac{\hbar}{m}P_{t}k_{+} & -\dfrac{\hbar}{m}P_{l}k_{z} \\
\dfrac{\hbar}{m}P_{l}k_{z} & \dfrac{\hbar}{m}P_{t}k_{-} & -\dfrac{E_{g}}{2} - \dfrac{\hbar^{2}k_{t}^{2}}{2m_{t}^{+}} - \dfrac{\hbar^{2}k_{z}^{2}}{2m_{l}^{+}} & 0 \\
\dfrac{\hbar}{m}P_{t}k_{+} &  -\dfrac{\hbar}{m}P_{l}k_{z} & 0 & -\dfrac{E_{g}}{2} - \dfrac{\hbar^{2}k_{t}^{2}}{2m_{t}^{+}} - \dfrac{\hbar^{2}k_{z}^{2}}{2m_{l}^{+}}
\end{pmatrix}.
\label{kp1}
\end{flalign}
In Eq.~\ref{kp1}, $ E_{g} $ is the energy gap, $ P_{l} $ and $ P_{t} $ are coupling constants, $ m_{l} $ and $ m_{t} $ denote the longitudinal and transverse effective masses, $ k_{t}^{2} = k_{x}^{2} + k_{y}^{2} $, and $ k_{\pm} = k_{x} \pm ik_{y} $. The material parameters for the lead chalcogenides, PbTe, PbS, and PbSe are collected in Table~\ref{table1}. 

From the bulk Hamiltonian, the corresponding variant for the quantum well is constructed on a finite-difference grid~\cite{sengupta2016numerical} by making the transformation $ k_{z} = -i\left(\partial/\partial\,z\right) $ in Eq.~\ref{kp1} for a quantized \textit{z}-axis aligned along the $ \left[111\right] $ direction. The dispersion of a $ 6.0\, nm $ wide PbTe QW along with the projected two-dimensional Brillouin zone is shown in Fig.~\ref{dsbz}. Note that Eq.~\ref{kp1} represents the \textit{L}-valley with axis along the $ \left[111\right] $ direction; however, there exist three other oblique valleys with identical symmetries. These oblique valleys have their axis tilted by $ \approx 70.5^{\circ} $ to the $ \left[111\right] $ direction, the Hamiltonian for which is written by a simple rotation of the coordinate system used in Eq.~\ref{kp1}. In the rotated system of coordinates used in Eq.~\ref{kp1} where the $ \left[111\right] $ vector serves as the \textit{z}-axis, the $ \hat{k}_{x} $, $ \hat{k}_{y} $, and $ \hat{k}_{z} $ terms must be replaced by performing the operation $ \mathbf{P}\mathbf{k} $, where $ \mathbf{P} $ is the transformation matrix and $ \mathbf{k} $ is the \textit{k}-vector triad. The transformation matrix, $ \mathbf{P} $ for one of the equivalent tilted valleys, say, $ \left[11\overline{1}\right] $ can be written noting that its axis can be made parallel to that of the longitudinal valley, $ \left[111\right] $, through an anti-clockwise rotation of $ \theta = cos^{-1}\left(1/3\right) $. Note that the orthonormal coordinate system in this case is $ \hat{x} = \left(i - j\right)/\sqrt{2} $, $ \hat{y} = -\left(i + j + 2k\right)/\sqrt{6} $, and $ \hat{z} = \left(i + j - k\right)/\sqrt{3} $. There exists, in addition to $ \left[11\overline{1}\right] $, another set of tilted equivalent valleys, namely, $\left[\overline{1}11\right]$ and $ \left[1\overline{1}1\right]$.
\begin{table}[h!]
\caption{4-band k.p parameters for PbS, PbSe, and PbTe. The mass terms (row entries 2-5) are expressed as a factor of the free electron mass, $ m_{0} = 9.1 \times 10^{-31}\, kg $. The parameters for PbS and PbSe are taken from Ref.~\citenum{kang1997electronic}. For PbTe parameters, we used Ref.~\citenum{bauer1992magneto}. Note that the band gap of PbTe drop to $ E_{g} = 0.19\, eV $ at $ T = 0\, K $. The current values are for $ T = 300\, K $.}
\centering
\label{table1}
\begin{tabular}{lccc}
\noalign{\smallskip} \hline \hline \noalign{\smallskip}
Parameters & PbS & PbSe & PbTe \\\hline
$E_{g}\,(eV)$ & 0.41 & 0.28 & 0.29 \\
$1/m^{-}_{t}$ & 1.9 & 4.3 & 16.667 \\
$1/m^{-}_{l}$  & 3.7 & 3.1 & 9.802 \\
$1/m^{+}_{t}$  & 2.7 & 8.7 & 9.80 \\
$1/m^{+}_{l}$  & 3.7 & 3.3 & 1.087 \\
P$_{t}^{2}/m_{0}\,(eV)$ & 1.5 & 1.5 & 2.975  \\
P$_{l}^{2}/m_{0}\,(eV)$ & 0.8 & 0.85 & 0.273 \\
\noalign{\smallskip} \hline \noalign{\smallskip}
\end{tabular}
\vspace{-0.4cm}
\end{table} 

\subsection*{Relation between Fermi energy and Rashba coupling parameter }
An outcome of the inclusion of the spin-orbit interaction Hamiltonians is that Fermi energy now depends on the strength of the Rashba spin-orbit coupling parameter. We quoted the results in the manuscript (Eqs.~\ref{fermif} and ~\ref{fermivec}), a quantitative calculation is given here. To prove this, we first derive the density of states (DOS) beginning with standard expression $ D\left(E\right) = \dfrac{1}{4\pi^{2}}\int\,d^{2}k\,\delta\left(E - \varepsilon\left(k\right)\right) $. Note that the Rashba-split conduction electron energy is of the form $ E = \alpha k^{2} \pm \lambda_{R}k $, where $ \alpha = \hbar^{2}/2m^{*} $. To evaluate the $ \delta\left(\cdot\right) $, we recall the identity $  \delta\left(g\left(k\right) \right) = \dfrac{\delta\left(k - k_{i}\right)}{\vert\,g^{'}\left(k_{i}\right)\vert} $, where $  k_{i} $ is the non-degenerate root of $ g\left(k\right) $. For our case, $ g\left(k\right) = E - \alpha k^{2} \mp \lambda_{R} k $, where the $ \mp $ differentiates the spin-up function from the spin-down version. The parameters $ \alpha $ and $ \beta $ are always positive. We retain the positive root for $ g\left(k\right) $ in each case giving $  k_{i} = \left(\sqrt{\lambda_{R}^{2} + 4\alpha E} \mp \lambda_{R}\right)/2\alpha $. Inserting all of this in the standard DOS equation gives
\begin{equation}
\begin{aligned}
D\left(E\right) &=  \dfrac{1}{4\pi^{2}}\int k\,dk\int_{0}^{2\pi} d\theta\dfrac{\delta\left(k - k_{i}\right)}{\vert\,-2\alpha k \mp \lambda_{R}\vert}, \\
&= \dfrac{1}{4\pi^{2}}\int_{0}^{2\pi} d\theta\dfrac{k_{i}}{\vert\,-2\alpha k_{i} \mp \lambda_{R} \vert} .
\label{dose}
\end{aligned}
\end{equation}
The integral evaluates to
\begin{equation}
D\left(E\right) = \dfrac{1}{2\pi}\dfrac{\sqrt{\lambda_{R}^{2} + 4\alpha E} \mp \lambda_{R}}{2\alpha\sqrt{\lambda_{R}^{2} + 4\alpha E}}.
\label{simpdos} 
\end{equation}
We have used $ k_{i} = \left(\sqrt{\lambda_{R}^{2} + 4\alpha E} \mp \lambda_{R}\right)/2\alpha $ in Eq.~\ref{simpdos}. The upper (lower) sign is for the spin-up (down) electrons. Further, rewriting Eq.~\ref{simpdos} as,
\begin{equation}
D\left(E\right) = \dfrac{1}{4\pi\alpha}\left[1 \mp \dfrac{\beta}{\sqrt{\beta^{2} + 4\alpha E}}\right],
\label{dosapp}
\end{equation}
it is easy to see that the spin-up branch has a lower DOS than its oppositely spin-polarized description. Finally, when $ \alpha k^{2} \gg \lambda_{R} k $ (or large energies), the DOS approaches the value $ \dfrac{1}{4\pi\alpha} = \dfrac{m^{*}}{2\pi\hbar^{2}} $. This is the standard DOS expression for a two-dimensional system without spin-degeneracy. We next calculate the electron density, $ n $, at Fermi energy (zero temperature), which by definition is $ n = \int_{0}^{\epsilon_{f}}D\left(E\right)dE $. Integrating after substituting for DOS from Eq.~\ref{simpdos} gives
\begin{equation}
\begin{aligned}
n &= \dfrac{1}{4\pi\alpha}\int_{0}^{\epsilon_{f}}\left[1 \mp \dfrac{\lambda_{R}}{\sqrt{\lambda_{R}^{2} + 4\alpha E}}\right]\,dE,\\
& = \dfrac{\epsilon_{f}}{4\pi\alpha} \mp \dfrac{\lambda_{R}}{8\pi\alpha^{2}}\left(\sqrt{\lambda_{R}^{2} + 4\alpha \epsilon_{f}} - \lambda_{R} \right).
\label{eldc}
\end{aligned}
\end{equation}
In Eq.~\ref{eldc}, the upper (lower) subscript of $ n $ is for the spin-up (down) branch. For a low effective mass as in PbTe, we can further simplify since $ \alpha = \dfrac{\hbar^{2}}{2m^{*}} \gg \lambda_{R} $; this reduces Eq.~\ref{eldc} to the following form
\begin{equation}
\dfrac{\epsilon_{f}}{4\pi\alpha} \mp \dfrac{\lambda_{R}}{4\pi\alpha^{3/2}}\sqrt{\epsilon_{f}} - n_{\pm} = 0.
\label{fermi1}
\end{equation}
Solving Eq.~\ref{fermi1} furnishes two values,
\begin{equation}
\epsilon_{f\pm} = \left[2\pi\alpha\left(\sqrt{\dfrac{\lambda_{R}^{2}}{16\pi^{2}\alpha^{3}} + \dfrac{n}{\pi\alpha}} \pm \dfrac{\lambda_{R}}{4\pi\alpha^{3/2}}\right)\right]^{2}.   
\label{fermif1}
\end{equation}
The upper (lower) sign is for spin-up (down) electrons. As a consistency check, setting $ \lambda_{R} \rightarrow 0 $, the Fermi energy in Eq.~\ref{fermif1} reduces to $ \epsilon_{f} = 4\pi\alpha n $ from which we recover the standard Fermi wave vector expression : $ k_{f} = 2\sqrt{\pi n} $. To obviate any confusion, the electron density $\left(n\right) $ is per spin branch. In addition, as remarked previously in the manuscript, Eq.~\ref{fermif1} shows the Fermi energy dependence on the strength of the Rashba spin-orbit coupling parameter. Lastly, the corresponding Fermi vector can be easily obtained: We simply use the relation, $ \epsilon_{f\pm} = \alpha k_{f}^{2} \pm \lambda_{R} k_{f} $ and solving for $ k_{f} $ yields
\begin{equation}
k_{f} = \dfrac{\sqrt{\lambda_{R}^{2} + 4\alpha\epsilon_{f\pm}} \mp \lambda_{R}}{2\alpha}.
\label{fwv}
\end{equation}
The upper (lower) sign in Eq.~\ref{fwv} is for the spin-up (down) ensemble. Note from Eq.~\ref{fwv} that the radius of the equi-energy circle for the spin-down band is larger than its spin-up branch.
 
\subsubsection*{Relaxation time from Born approximation}
We noted above the possibility of calculating the relaxation time for carriers scattered by surface impurities within the Born approximation. Assuming that the primary source that impedes electronic motion is impurity-driven (phonon modes and their coupling to the electronic ensemble are suppressed), the imaginary part of the retarded self-energy $ \left(\Sigma\right) $ for such an ensemble of conduction electrons allows us to estimate the scattering time $\left(\tau_{p}\right) $ through the relation, $  1/\tau_{p} = \left(2/\hbar\right)Im\Sigma $. The retarded self-energy in the Born approximation (SCBA) is expressed as a pair of equations~
\begin{equation}
\begin{aligned}
\label{scba1}
G_{ks}\left(\epsilon\right) = \dfrac{1}{\epsilon - \epsilon_{ks} - \Sigma\left(\epsilon\right)};
\Sigma\left(\epsilon\right) = n_{i}v_{i}^{2}\int\,\dfrac{d^{2}k}{4\pi^{2}}G_{ks}\left(\epsilon\right),
\end{aligned}
\end{equation}
where $ n_{i} $ and $ v_{i} $ denote the density and strength of impurities, respectively and $ G_{ks}\left(\epsilon\right) $ is the $ 2 \times 2 $ retarded Green's function diagonal with respect to the band index \textit{s} ($\langle\,s\vert\,G_{k}\left(\epsilon\right)\vert\,s^{'}\rangle = \delta_{ss^{'}}G_{ks}\left(\epsilon\right) $). The retarded self-energy, $ \Sigma $, in SCBA averaged over impurity distributions is also diagonal with respect to the band index \textit{s} and independent of \textbf{\textit{k}}.

The retarded Green's function corresponding to the Hamiltonian in Eq.~\ref{hrs} is
\begin{equation}
G^{R} = \begin{pmatrix}
E - \alpha k^{2} + i0^{+} &  -\lambda_{R}\left(k_{y} + ik_{x}\right) \\
-\lambda_{R}\left(k_{y} - ik_{x}\right) & E - \alpha k^{2} + i0^{+}
\end{pmatrix}^{-1}.
\label{ginv}
\end{equation}
The retarded Green's function from Eq.~\ref{ginv} when inserted in Eq.~\ref{scba1} and integrating the diagonal elements of the $ 2 \times 2 $ matrix in two-dimensional space, the self energy $\left(\Sigma\right) $ term has the form
\begin{equation}
\Sigma = \dfrac{n_{i}v_{i}^{2}}{8\pi^{2}} \int d^{2}k \biggl[\dfrac{1}{E - \alpha k^{2} - \lambda_{R} k + i0^{+}} + \dfrac{1}{E - \alpha k^{2} + \lambda_{R} k + i0^{+}} \biggr].
\label{rgf1}
\end{equation}
Employing the standard relation $ \dfrac{1}{x \pm i\delta} =  \mathbb{P}\dfrac{1}{x} \mp i\pi\delta\left(x\right) $, we arrive at
\begin{align}
Im\Sigma\left(E\right) &= \dfrac{n_{i}v_{i}^{2}}{8\pi}\int d^{2}k\left(\delta_{1} + \delta_{2}\right) ,
\label{rgf2}
\end{align}
where $ \delta_{1} =\delta\left(E - \alpha k^{2} - \lambda_{R} k\right) $ and $ \delta_{2} =\delta\left(E - \alpha k^{2} + \lambda_{R} k\right) $. The integration is performed by changing over in to energy space and approximating the energy differential $ dE = 2\alpha k dk $; at the Fermi surface for spin-up and spin-down branches, the integral evaluates to $ m^{*}n_{i}v_{i}^{2}/2\hbar^{2} $. Note that we have replaced $ \alpha $ with $ \hbar^{2}/2m^{*} $. For materials such as PbTe, where a low effective mass is the significant contributor to the total Hamiltonian, we can ignore the Rashba spin part, and assuming that localized impurity scattering is dominant, the relaxation time $ \left(1/\tau_{p} = \left(2/\hbar\right)Im\Sigma\right) $ is energy-independent. As a caveat~\cite{0953-8984-29-40-405701}, the relaxation time obtained from the imaginary part of the self-energy is not true for scattering events with a small angular spread. For a fuller discussion, see, for example, Chap. 8, Ref.~\citenum{mahan2013many}.

\noindent \textbf{Data availability} : All data analyzed in preparation of this work are included here.



\section*{Author contributions statement}
P.S. conceived the work. P.S. and J.S. analyzed the results. All authors contributed in writing the manuscript. 

\section*{Competing financial interests}
The authors declare no competing financial interests.

\end{document}